\RequirePackage{lineno}
\documentclass[twocolumn,showkeywords,superscriptaddress,preprintnumbers,amsmath,amssymb,prc]{revtex4}  
\usepackage{graphicx}
\usepackage{dcolumn}
\usepackage{bm}
\usepackage{float}
\usepackage{color}
\usepackage{CJK}
\usepackage[colorlinks,linkcolor=blue,urlcolor=blue,citecolor=blue]{hyperref}
\usepackage{isotope}
\usepackage[normalem]{ulem}
\renewcommand{\sout}{\bgroup \color{red} \ULdepth=-0.5ex \ULset}

\usepackage{amsmath}
\usepackage{cases}
\usepackage{amssymb}
\usepackage{eqnalign}

\begin{document}
\begin{CJK*}{UTF8}{gbsn}


\title{Nuclear giant quadruple resonance within transport approach and its constraint on nucleon effective mass}

\author{Yi-Dan Song(宋一丹)}
\affiliation{Key Laboratory of Nuclear Physics and Ion-beam Application~(MOE), Fudan University, Shanghai $200433$, China}

\author{Rui Wang(王睿)}
\email{wangrui@sinap.ac.cn}
\affiliation{Shanghai Institute of Applied Physics, Chinese Academy of Sciences, Shanghai $201800$, China}

\author{Zhen Zhang(张振)}
\affiliation{Sino-French Institute of Nuclear Engineering and Technology, Sun Yat-sen University, Zhuhai $519082$, China}

\author{Yu-Gang Ma(马余刚)}
\email{mayugang@fudan.edu.cn}
\affiliation{Key Laboratory of Nuclear Physics and Ion-beam Application~(MOE),  Fudan University, Shanghai $200433$, China}
\affiliation{Shanghai Institute of Applied Physics, Chinese Academy of Sciences, Shanghai $201800$, China}

\date{\today}

\begin{abstract}
{
We study the nuclear iso-scalar giant quadruple resonance~(ISGQR) based on the Boltzmann-Uehling-Uhlenbeck~(BUU) transport equation.
The mean-field part of the BUU equation is described by the Skyrme nucleon-nucleon effective interaction, and its collision term, which embodies the two-particle-two-hole ($2$p-$2$h) correlation, is implemented through the stochastic approach.
We find that the width of ISGQR for heavy nuclei is exhausted dominated by collisional damping, which is incorporated into the BUU equation through its collision term, and it can be well reproduced through employing a proper in-medium nucleon-nucleon cross section.
Based on further Vlasov and BUU calculations with a number of representative Skyrme interactions, the iso-scalar nucleon effective mass at saturation density is extracted respectively as $m^{*}_{s,0}/m$ $=$ $0.83\pm0.04$ and $m^{*}_{s,0}/m$ $=$ $0.82\pm0.03$ from the measured excitation energy $E_x$ of the ISGQR of $\isotope[208]{Pb}$. 
The small discrepancy between the two constraints indicates the negligible role of $2$p-$2$h correlation in constraining $m_{s,0}^*$ with the ISGQR excitation energy.
}

\end{abstract}

\maketitle

\section{Introduction}\label{1}
Nuclear giant resonances are a global feature of nuclei that arises from the collective motion of their constituent nucleons \cite{BerRMP55,PitPLB36,BerNPA354,YouPRC23,KnoPRL46,BueJPC45,YouPRC69,GuptaPRC97,MonPRL100,HeWB,HuangBS,Wang1,Wang2}.
Following the iso-vector giant dipole resonance~(IVGDR)~\cite{BerRMP55}, the iso-scalar giant quadruple resonance~(ISGQR), discovered in the early $1970$s~\cite{PitPLB36}, was the second fundamental mode of nuclear giant resonances.
Systematic studies on giant resonance for different nuclear masses have been carried out by, ($d$, $d'$) or ($\alpha$, $\alpha'$) reactions~\cite{BerNPA354,YouPRC23,KnoPRL46,BueJPC45,YouPRC69,GuptaPRC97,MonPRL100}, which preferentially excites iso-scalar transitions due to their zero isospin nature.
One of the important features of the nuclear ISGQR is that its excitation energy $E_x$ is related to the nucleon effective mass in symmetric nuclear matter, or iso-scalar nucleon effective mass, $m_s^*(\rho)$~\cite{BlaPR64}.
It is well known that the strong exchange effects of nuclear interaction lead to a momentum dependence of nucleon potential in nuclear medium~\cite{WirPRC38}.
The nucleon effective mass is usually employed to characterize this momentum-dependence of the single-nucleon potential, and it plays an important role in the dynamics of heavy-ion collisions~\cite{LBAPR464,ZhangF_NST,LiBA_NST,WeiGF_NST}.
Apart from this, the isospin splitting of the nucleon effective masses, which is related to the $m_s^*$ and the iso-vector nucleon effective mass $m_v^*$, has strong influences on various quantities or processes, e.g., the properties of mirror nuclei~\cite{NolARNS19}, transport properties of asymmetric nuclear matter~\cite{LBAPRC72,LBANPA735}, neutrino emission in neutron stars~\cite{BalPRC89}.
Therefore, how to constrain accurately the nucleon effective mass has been a hot topic in nuclear physics, and one may refer to Ref.~\cite{LBAPPNP99} for a recent review.

Since the excitation energy $E_x$ of the nuclear ISGQR is an effective probe of the $m_{s}^*$, several methods have been employed to calculate the $E_x$ of the ISGQR and consequently constrain the $m_{s}^*$, e.g., the (quasi-)random-phase approximation~(RPA) approach~\cite{BohPR51,RocPRC87,YosPRC88,ZZPRC93,YukPRC97,BonPRC98,ZZCPC45}, the macroscopic Langevin equation~\cite{SJPRC101}, and dynamical approaches like transport models~\cite{KHYPRC95,XJPRC102}.
One of the advantages of the transport model is that the beyond mean-field effects can be easily included via the collision term.
The Boltzmann-Uehling-Uhlenbeck~(BUU) equation~\cite{BerPR160} is one of the main transport models that has been widely applied to the study of heavy-ion collisions.
On the one hand, with the absence of the collision term, the BUU equation reduces to the Vlasov equation, which can be regarded as a semi-classical approximation of the mean-field level time-dependent Hartree-Fock (density functional) approach~\cite{NakRMP88,SimPPNP103,StePPNP104}.
On the other hand, the nucleon-nucleon scatterings in the BUU equation embodies beyond-mean-field correlations, or the two-particle-two-hole ($2$p-$2$h) correlation, and it incorporates two-body dissipation into the evolution.
It has been shown that the nucleon-nucleon scatterings are essential to describe the experimental full width at half maximum $\Gamma$ of IVGDR for heavy nuclei~\cite{WRPLB807}.

Although there already exist several works that aim to study the ISGQR and its constraint on the $m_{s}^*$ using the BUU transport equation~\cite{KHYPRC95,XJPRC102}, they suffer from the problems of unstable nuclear ground-state evolution and inaccurate treatment of the collision term, especially the Pauli blocking, which has been shown to have significant effect on the width $\Gamma$ of nuclear giant resonances~\cite{WRPLB807,WRFiP8}.
In the present work, to study the ISGQR of finite nuclei, we employ a lattice Hamiltonian method~\cite{LenPRC39} to solve the BUU equation, and its nucleon-nucleon collision term is implemented through a full-ensemble stochastic collision approach~\cite{WRPRC99,WRPLB807,WRFiP8}.
The ground-state of a nucleus is obtained by varying the total energy with respect to the nucleon density distribution based on the same Hamiltonian that governs the equations of motion of nucleons.
Besides that, the present framework of solving the BUU equation is implemented numerically through GPU~(Graphics Processing Unit) high-performance parallel computing~\cite{Rue2013}, which enables us to employ huge amount of test particles when solving the BUU equation, and increases the numerical efficiency and accuracy profoundly.
The above features help to improve the ground-state stability and the accuracy of the collision term significantly, and make it possible to solve the BUU equation precisely.
Therefore, one might describe through the BUU equation the time evolution of the ISGQR and obtain its strength function $S(E)$, and subsequently its excitation energy $E_x$ and width $\Gamma$.

The present article is organized as follows.
In Sec.~\ref{2}, we introduce the basic concept of solving the BUU equation, and the necessary formalism of nucleon effective masses in Skyrme-Hartree-Fock (SHF) approach, as well as how to study nuclear collective motions in the transport approach.
In Sec.~\ref{3}, we study the ISGQR of \isotope[208]{Pb} and its strength function within the BUU equation, and present a correlation between its excitation energy $E_x$ and iso-scalar nucleon effective mass $m_s^*$ through the BUU calculation from various Skyrme interactions, based on which we constrain the $m_{s}^*$.
Finally, we summarize the present work in Sec.~\ref{4}.

\section{Method}\label{2}

In the present study, the ISGQR of finite nuclei is studied within the framework of the BUU transport equation.
In this section, we briefly present how we solve the BUU equation effectively, i.e., the lattice Hamiltonian method adopted for the mean-field evolution and the stochastic method employed for the collision term, and how to describe the nuclear collective motions with Wigner function $f(\vec{r},\vec{p})$.
Since we employ the Skyrme effective nucleon-nucleon interaction to describe the mean-field part of the BUU equation, we also introduce several basic features of the Skyrme interaction and nucleon effective masses within the Skyrme interaction.

\subsection{Boltzmann-Uehling-Uhlenbeck transport equation}

The BUU transport equation describes the time evolution of the Wigner function $f(\vec{r},\vec{p})$.
If we consider a momentum-dependent potential $U(\vec{r},\vec{p})$, the BUU equation can be written as,
\begin{equation}\label{E:BUU}
\frac{\partial{f}}{\partial{t}}+\frac{\vec{p}}{E}\cdot\triangledown_{\vec{r}}f+\triangledown_{\vec{p}}U(\vec{r},\vec{p})\cdot\triangledown_{\vec{r}}f-\triangledown_{\vec{r}}U(\vec{r},\vec{p})\cdot\triangledown_{\vec{p}}f = I_{c}.
\end{equation}
In the above equation, the left hand side describes the time evolution of the $f(\vec{r},\vec{p})$ in the mean-field $U(\vec{r},\vec{p})$, while the collision integral $I_c$ is responsible for part of the beyond mean-field many-body correlation.
The general form of $I_c$ is
\begin{equation}\label{E:Ic}
\begin{split}
I_{c} = &-g\int\frac{d^{3}p_{2}}{(2\pi\hbar)^{3}}\frac{d^{3}p_{3}}{(2\pi\hbar)^{3}}\frac{d^{3}p_{4}}{(2\pi\hbar)^{3}}\\
     &\times|\mathcal{M}_{12\to34}|^{2}(2\pi)^{4}\delta^{4}(p_{1}+p_{2}-p_{3}-p_{4})\\
     &\times\big[ff_{2}(1-f_{3})(1-f_{4})-f_{3}f_{4}(1-f)(1-f_{2})\big],
\end{split} 
\end{equation}
where $f_i$ is short for $f(\vec{r},\vec{p}_i)$, $g$ is a degeneracy factor, and $\mathcal{M}_{12\to34}$ is the in-medium transition matrix element.
The $1 - f$ in $I_c$ is added to take  the Pauli principle into account due to nucleons’ Fermion statistics.

The lattice Hamiltonian method~\cite{LenPRC39} is adopted to solve the mean-field evolution of the BUU equation.
We first mimic $f$ by a large number of test nucleons~\cite{WonPRC25}, 
\begin{equation}\label{E:f_TP}
f(\vec{r},\vec{p},t) = \frac{1}{g}\frac{(2\pi\hbar)^{3}}{N_{E}}\sum^{AN_{E}}_{i}S[\vec{r}_{i}(t)-\vec{r}~]\delta[\vec{p}_{i}(t)-\vec{p}~],
\end{equation}
with $A$ to be the mass number of the system and $N_E$ the number of parallel ensembles~(or the number of test particles in some literature).
A form factor $S$ in coordinate space is introduced to modify the relation between test nucleons and $f$.
Based on the $f$ in Eq.~(\ref{E:f_TP}), we then calculate the total Hamiltonian $H$ approximated by the lattice Hamiltonian $H_L$,
\begin{equation}\label{eq.26}
H = \int\mathcal{H}(\vec{r})d\vec{r}\thickapprox l_xl_yl_z\sum_{\alpha}\mathcal{H}(\vec{r}_{\alpha})\equiv H_L,
\end{equation}
to obtain the equations of motion of test nucleons and subsequently other physical quantities.
In the above formula, $\vec{r}_{\alpha}$ represents certain lattice site, and $l_x$, $l_y$, $l_z$ are lattice spacing.

We adopt the stochastic collision approach~\cite{DanNPA533,XZPRC75} to deal with the collision term $I_c$.
The scattering probability of a scattering event, involving test nucleons $i$ and $j$ in a time interval $\Delta t$, can be calculated directly through the collision term by substituting the $f$ in Eq.~(\ref{E:f_TP}) into Eq.~(\ref{E:Ic}).
It reads 
\begin{equation}\label{E:Pij}
P_{ij} = \upsilon_{rel}\sigma^{*}_{\rm NN}S(\vec{r}_{i}-\vec{r}_{\alpha})S(\vec{r}_{j}-\vec{r}_{\alpha})l_{x}l_{y}l_{z}\Delta t,
\end{equation}
where $\upsilon_{rel}$ is the relative velocity between the two test nucleons, and $\sigma^{*}_{\rm NN}$ is the in-medium nucleon-nucleon cross section.
Note its difference with the probability in normal stochastic approach~\cite{XZPRC75}, due to the existence of the form factor $S$ in the present study.
If the collision of the $i$th and $j$th test nucleons happens, we calculate a Pauli blocking factor $\big[1-f(\vec{r}_{\alpha},\vec{p}_i')\big]\big[1-f(\vec{r}_{\alpha},\vec{p}_j')\big]$ according to their final state momentum $\vec{p}_i'$ and $\vec{p}_j'$, to determine whether the collision is blocked by the Pauli principle.

We have omitted the isospin degree of freedom in the above context for clarity.
The present lattice BUU framework of solving the BUU equation has been applied successfully to the giant resonances of heavy nuclei~\cite{WRPRC99,WRPLB807}.
One can find more detailed descriptions about the present framework, e.g., the equations of motion of test nucleons, in Ref.~\cite{WRFiP8}.

\subsection{Nucleon effective mass in Skyrme-Hartree-Fock approach}

As mentioned above, we employ the SHF approach~\cite{VauPRC5,ChaNPA627,ChaNPA635} to describe the mean-field in the BUU equation.
The standard parametrization form of the Skyrme interaction~\cite{ChaNPA635,ChaNPA627} is,
\begin{equation}\label{E:Sky}
\begin{split}
\upsilon_{12}& = t_{0}(1+x_{0}{\hat{P}_{\sigma}})\delta(\vec{r}_{1}-{\vec{r}_{2}})\\
						&+\frac{1}{6}t_{3}(1+x_{3}{\hat{P}_{\sigma}})\rho^{\alpha}(\vec{R})\delta(\vec{r}_{1}-{\vec{r}_{2}})\\
					        &+t_{1}(1+x_{1}{\hat{P}_{\sigma}})\frac{1}{2}\Big[\delta(\vec{r}_{1}-{\vec{r}_{2}}){\hat{\vec{k}} ^{'2}}+\hat{\vec{k}}^{2}\delta(\vec{r}_{1}-{\vec{r}_{2}})\Big]\\
					        &+t_{2}(1+x_{2}{\hat{P}_{\sigma}})\hat{\vec{k}}^{'}\cdot\delta(\vec{r}_{1}-{\vec{r}_{2}})\hat{\vec{k}}\\
					        &+iW_{0}(\hat{\vec{\sigma}}_{1}+\hat{\vec{\sigma}}_{2})\cdot\hat{\vec{k}}^{'}\times\delta(\vec{r}_{1}-{\vec{r}_{2}})\hat{\vec{k}},
\end{split}     		
\end{equation}
with the $\vec{R} = \frac{1}{2}(\vec{r}_{1}+{\vec{r}_{2}})$.
The relative momentum operators $\hat{\vec{k}}=(\hat{\vec{\nabla}}_1 - \hat{\vec{\nabla}}_2)/2i$ and $\hat{\vec{k}}^{'}=-(\hat{\vec{\nabla}}_1 - \hat{\vec{\nabla}}_2)/2i$ act  on the right and left, respectively.
The $\hat{\vec{\sigma}}_{i}$ and $\hat{\vec{P}}_{\sigma}$ are the Pauli spin matrices and the spin exchange operator, respectively. 
The $t_{0}$ $-$ $t_{3}$, $x_{0}$ $-$ $x_{3}$, $\alpha$ are Skyrme parameters.
In Eq.~(\ref{E:Sky}), the first term is the local density term, and the second term is the density-dependent term, which is used to describe effectively the three-body interaction between nucleons.
The third and fourth terms are momentum dependent terms, and the fifth term is spin-orbit coupling term which is omitted in the present study due to the assumption of spin saturation.

We can include in Eq.~(\ref{E:Sky}) additional high-order momentum dependent terms, to form the Skyrme pseudopotential~\cite{CarPRC78,RaiPRC83}.
The Skyrme pseudopotential with next-to-next-to-next-to-leading order~(N$3$LO) momentum dependent terms can reproduce the empirical nuclear optical potential up to about 1 GeV in kinetic energy~\cite{WRPRC98}.

In the SHF approach~\cite{VauPRC5,ChaNPA627,ChaNPA635}, the nucleon effective mass $m^*_{q}/m$ for a nucleon $q$~($n$ or $p$) in asymmetric nuclear matter with density $\rho$ can be obtained as follow,
\begin{equation}\label{eq.28}
\begin{split}
\frac{\hbar^2}{2m^*_{q}(\rho,\delta)}&=\frac{\hbar^2}{2m}+\frac{1}{4}t_{1}\Big[(1+\frac{1}{2}x_{1})\rho-(\frac{1}{2}+x_{1})\rho_{q}\Big]\\
						     &+\frac{1}{4}t_{2}\Big[(1+\frac{1}{2}x_{2})\rho+(\frac{1}{2}+x_{2})\rho_{q}\Big],
\end{split}     		
\end{equation}
with isospin asymmetry $\delta = (\rho_{n}-\rho_{p})/(\rho_{n}+\rho_{p})$. 
When setting $\rho_{q} = \rho/2$, we obtain the iso-scalar nucleon effective mass
\begin{equation}\label{eq.29}
\frac{\hbar^2}{2m^*_s(\rho)} = \frac{\hbar^2}{2m}+\frac{3}{16}t_{1}\rho+\frac{1}{16}t_{2}(5+4x_{2})\rho.
\end{equation}
It's expression in the Skyrme pseudopotential can be obtained straightforwardly.

\subsection{Nuclear collective motions in the Wigner representation}

One of the key quantities when studying the collective motions is the expectation value of the excitation operator $\hat{Q}$.
In the Wigner representation~\cite{CarRMP55,BonPRL71}, i.e., the variable in the equation of motion has been transformed from wave function or density matrix to $f(\vec{r},\vec{p})$, we should express the expectation $\langle{\hat{Q}}\rangle$ in terms of $f(\vec{r},\vec{p})$.
The $\hat{Q}$ is usually considered as a one-body operator, which means that it can be expressed as the sum of single-nucleon operator ${\hat{q}_{i}}$ acting on each nucleon. 
The expectation of $\hat{Q}$ can then be expressed as
\begin{eqnarray}\label{E:Q1}
\langle{\hat{Q}}\rangle \equiv&\langle\phi|{\hat{Q}}|\phi\rangle={\sum^{A}_{i}\langle\phi|{\hat{q}_{i}}|\phi\rangle}\notag\\
		= &\sum^{A}_{i}\int{\langle\phi|\vec{r}_{1}...\vec{r}_{N}\rangle\langle\vec{r}_{1}...\vec{r}_{N}|{\hat{q}_{i}}|{\vec{r}_{1}}'...\vec{r}_{N}'}\rangle\notag\\ 
		&\times\langle{\vec{r}_{1}'}...{\vec{r}_{N}'}|\phi\rangle{d^{3}\vec{r}_{1}}...{d^{3}\vec{r}_{N}}{d^{3}\vec{r}_{1}'}...{d^{3}\vec{r}_{N}'}\notag\\
		= &\int{\rho(\vec{r}_{i}',\vec{r}_{i})\langle\vec{r}_{i}|{\hat{q}_{i}}|{\vec{r}_{i}'}\rangle d^{3}\vec{r}_{i}{d^{3}\vec{r}_{i}'}},
\end{eqnarray}
where $|\vec{r}_i\rangle$ and $|\vec{r}'_i\rangle$ are coordinate eigen-states, and $\rho({\vec{r}_{i}'},\vec{r}_{i})$ is the one-body density matrix.
By transforming the relative coordinate $\vec{r}_i-\vec{r}_i'$ in Eq.~(\ref{E:Q1}) to momentum space, we obtain $\langle{\hat{Q}}\rangle$ in terms of the Wigner function $f(\vec{r},\vec{p})$, i.e.,
\begin{equation}\label{E:Q2}
\langle{\hat{Q}}\rangle = \int{f(\vec{r},\vec{p})q(\vec{r},\vec{p})d^{3}\vec{r}d^{3}\vec{p}},
\end{equation}
where the $q(\vec{r},\vec{p})$ is the Wigner transform of $\langle\vec{r}_{i}|{\hat{q}_{i}}|{\vec{r}_{i}'}\rangle$ and $\vec{r}$ $=$ $(\vec{r}_i-\vec{r}_i')/2$. 
Therefore, the expectation of $\hat{Q}$ can be calculated within the framework of BUU equation.
The specific form of the excitation operator $\hat{Q}$ for iso-scalar quadruple mode will be given in Sec.~\ref{S:31}.

\section{Results and discussion}\label{3}

\subsection{Strength function of ISGQR}\label{S:31}

In transport approach, an excited nucleus can be generated by changing the $f(\vec{r},\vec{p})$ of a ground state nucleus according to the form of the excitation operator.
This is equivalent to changing the initial coordinates and momenta of test nucleons if we adopt the test particle ansatz.
For the coordinates $\vec{r}_i$ and momenta $\vec{p}_i$ of test nucleons of a ground state nucleus, since we deals with a semi-classical transport equation, they are obtained based on the Thomas-Fermi approach.
In the Thomas-Fermi approach, the radial density distribution $\rho_{\tau}(r)$ of a ground state nucleus is obtained by varying the total Hamiltonian with respect to it.
Since the same Hamiltonian also governs the evolution of the test nucleons, a very stable ground state evolution can be obtained in the transport calculation.
The initial coordinates $\vec{r}_i$ of test nucleons are generated according to the obtained radial density distribution, while their initial momenta $\vec{p}_i$ are generated from a zero-temperature Fermi distribution with the Fermi momentum
given as $p^{\rm F}_{\tau}$ $=$ $\hbar\big[3\pi^2\rho_{\tau}(r)\big]^{1/3}$.

For the ISGQR, the excitation operator ${{\hat{Q}}_{\rm ISQ}}$ can be written as the sum of single-nucleon operator ${{\hat{q}}_{\rm ISQ}}$,
\begin{equation}
{\hat{Q}_{\rm ISQ}} = \sum^A_{\rm i=1}\hat{q}_{\rm ISQ} = {\sum^{A}_{i=1}}\frac{1}{A}\sqrt{\frac{5}{16\pi}}(3{\hat{z}_{i}}^{2}-\hat{r}_{i}^{2}),
\end{equation}
where $\hat{r}_{i}$ and $\hat{z}_{i}$ are the coordinate operator of the $i$-th nucleon and its $3$-rd component, respectively, and $A$ is the mass number of the nucleus.
For ${{\hat{q}}_{\rm ISQ}}$, its corresponding ${q_{\rm ISQ}}(\vec{r},\vec{p})$ in Eq.~(\ref{E:Q2}) is expressed as
\begin{equation}\label{eq.32}
{q_{\rm ISQ}}(\vec{r},\vec{p}) = \frac{1}{A}\sqrt{\frac{5}{16\pi}}(3z^2-\vec{r}\,^2).
\end{equation}
Note here the coordinate operators have been replaced by their eigenvalues.

To obtain the iso-scalar quadruple mode, we can either change the initial $\vec{r}_i$ and $\vec{p}_i$ of test nucleons according to
\begin{equation}\label{E:ex1}
\vec{p}_{i}\rightarrow\left\{
\begin{aligned}
{p_{xi}}+2{\lambda}\sqrt{\frac{5}{16\pi}}\frac{x_{i}}{A}\\
{p_{yi}}+2{\lambda}\sqrt{\frac{5}{16\pi}}\frac{y_{i}}{A}\\
{p_{zi}}-4{\lambda}\sqrt{\frac{5}{16\pi}}\frac{z_{i}}{A},\\
\end{aligned}
\right.
\end{equation}
or according to
\begin{equation}\label{E:ex2}
\vec{r}_{i}\rightarrow\left\{
\begin{aligned}
x_i - x_i\lambda\\
y_i - y_i\lambda\\
z_i + 2z_i\lambda
\end{aligned}
\right.
\qquad 
\vec{p}_{i}\rightarrow\left\{
\begin{aligned}
p_{xi} + 2p_{xi}\lambda\\
p_{yi} + 2p_{yi}\lambda\\
p_{zi} - p_{zi}\lambda,
\end{aligned}
\right.
\end{equation}
with $\lambda$ to be a small quantity. 
The former scenario corresponds to exciting the nucleus at $t$ $=$ $t_0$, i.e., $\hat{H}_{\rm ex}$ $\propto$ $\lambda\hat{Q}_{\rm ISQ}\delta(t - t_0)$, which is a common practice within the linear response theory~\cite{Fet1971}.
The latter scenario corresponds to starting the nucleus at the ground state with the presence of the excitation operator~\cite{PatNPA703}, and then remove the excitation at $t$ $=$ $t_0$, i.e., $\hat{H}_{\rm ex}$ $\propto$ $\lambda\hat{Q}_{\rm ISQ}\theta(t_0 - t)$.
We will examine both of these two scenarios when calculating the ISGQR.

After exciting the nucleus to a certain mode, we calculate the time evolution of the expectation of the excitation operator $\langle\hat{Q}\rangle(t)$ within the BUU equation through Eq.~(\ref{E:Q2}), and subsequently obtain the strength function $S(E)$ for that mode.
In the following we calculate the ISGQR of \isotope[208]{Pb} with the Skyrme interaction SkM*~\cite{BarNPA386}.
We consider two cases, i.e., the pure Vlasov calculation, in which only the mean-field, i.e., one-particle-one-hole ($1$p-$1$h) correlation is taken into account, and the full BUU calculation, with the nucleon-nucleon collision term reflects effectively additional $2$p-$2$h correlation.
This nucleon-nucleon collision term is responsible for the collisional damping of the nuclear collective motions, which is indispensable for describing the width of the latter~\cite{WRPLB807}.
It has been shown that a large in-medium reduction of nucleon-nucleon cross section is essential to reproduce the experimental width of IVGDR of \isotope[208]{Pb}~\cite{WRPLB807}.
In the full BUU calculation, for simplicity, we employ a constant in-medium correction, i.e., $\sigma_{\rm NN}^*$ $=$ $0.60\sigma_{\rm NN}^{\rm free}$.
The free nucleon-nucleon cross section $\sigma_{\rm NN}^{\rm free}$ is parameterized based on the experimental nucleon-nucleon scattering data as in Ref.~\cite{CugNIMB111}.
This in-medium nucleon-nucleon cross section can reproduce the weighted average of the experimental width $\Gamma$ $=$ $3.0\pm0.1$~\cite{RocPRC87} of ISGQR of \isotope[208]{Pb}.
In the present study, we choose lattice spacing $l_x$ $=$ $l_y$ $=$ $l_z$ $=$ $0.5~\rm fm$, time step $\Delta{t} =\ 0.5\,\text{fm}/c$, and $N_{E}$ $=$ $10000$, to ensure the convergence of the numerical results.

\begin{figure}[!h]
\includegraphics
[width=7.5cm]{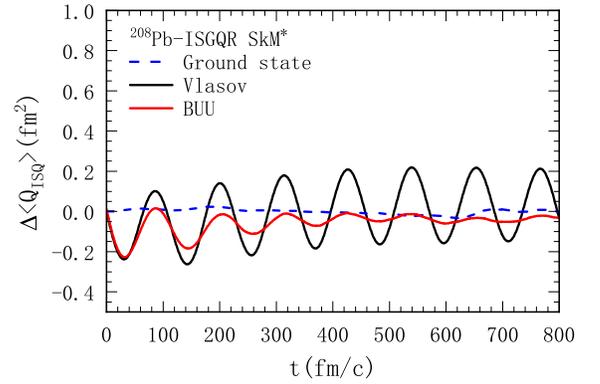}
\caption{The time evolution of the $\Delta\langle\hat{Q}_{\rm ISQ}\rangle$ of $\isotope[208]{Pb}$ in the ground-state and after a perturbation excitation of ${\hat{H}}_{ex}(t)$ $\propto$ ${\lambda}{\hat{Q}}_{\rm ISQ}{\delta(t - t_0)}$ with $\lambda = 0.1\ {\rm GeV}\cdot fm^{-1}/c$ based on Eq.~(\ref{E:ex1}). 
The results correspond to the ground-state, the pure Vlasov and the full BUU calculation, respectively, which are calculated with the Skyrme interaction SkM*. 
} \label{F:Q1}
\end{figure}

In Fig.~\ref{F:Q1}, we show the time evolution of the $\Delta\langle{\hat{Q}}_{\rm ISQ}\rangle$ through Eq.~(\ref{E:ex1}) with the Skyrme interaction SKM$^*$.
The black and red solid lines represent the $\Delta\langle{\hat{Q}}_{\rm ISQ}\rangle$ for pure Vlasov and full BUU calculations, respectively.
In both cases, the expectation of the excited nucleus $\langle\phi|{\hat{Q}}_{\rm ISQ}|\phi\rangle$ has been subtracted by the that of the ground state nucleus $\langle 0|{\hat{Q}}_{\rm ISQ}|0\rangle$.
The latter is expected to be zero, and it is shown in Fig.~\ref{F:Q1} with the blue dash line.
We notice from the figure that in the Vlasov calculation where only the mean-field ($1$p-$1$h) is included, the $\Delta\langle{\hat{Q}}_{\rm ISQ}\rangle$ shows a regular oscillation, with the amplitude of the oscillation almost unchanged.
Meanwhile, in the full BUU calculation where nucleon-nucleon scatterings are also included, the oscillation of the $\Delta\langle{\hat{Q}}_{\rm ISQ}\rangle$ damps very quickly and return to zero eventually.
The above feature indicates that the collisional damping, which is a two-body dissipation, dominates the ISGQR width of heavy nuclei, while the Landau damping, the one-body dissipation that relates only to the mean-field, has negligible effect on it.
This is different from the IVGDR case where the Landau damping is responsible for almost half of the width~\cite{WRPLB807}.
Such a feature suggests the ISGQR to be an ideal site for constraining the in-medium nucleon-nucleon cross section.

\begin{figure}[htbp]
\includegraphics
[width=7.5cm]{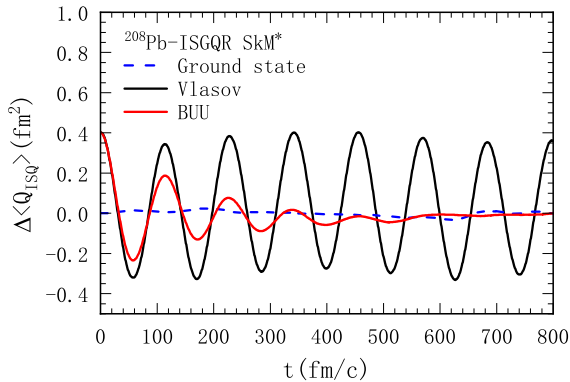}
\caption{
Same as Fig.~\ref{F:Q1} but for perturbation excitation of 
${\hat{H}}_{ex}(t)$ $\propto$ ${\lambda}{\hat{Q}_{\rm ISQ}}{\theta(t_0 - t)}$ with $\lambda=0.01\ {\rm GeV}\cdot fm^{-1}/c$.
} \label{F:Q2}
\end{figure}

Similarly, we show the time evolution of the $\Delta\langle{\hat{Q}}_{\rm ISQ}\rangle$ of the $\isotope[208]{Pb}$ after a perturbative iso-scalar quadruple excitation through Eq.~(\ref{E:ex2}).
Note that since in this case, the nucleus starts at the ground state with the presence of the excitation operator $\hat{Q}_{\rm ISQ}$, the initial $\Delta\langle\hat{Q}_{\rm ISQ}\rangle$ for the pure Vlasov and full BUU calculations, shown as solid line in the figure, have finite values.
Similar features with Fig.~\ref{F:Q1} about the damping of the $\Delta\langle{\hat{Q}}_{\rm ISQ}\rangle(t)$ for Vlasov and BUU calculations can be observed in Fig.~\ref{F:Q2}.
It is obviously seen that the result of the Vlasov calculation has similar oscillation with Fig.~\ref{F:Q1}, but more regular.
One explanation to this phenomenon is that if one excites the nucleus through Eq.~(\ref{E:ex1}), it is easier for the test nucleons at the surface of the nucleus along $x$ and $y$ axis to escape, whereas in the excitation scenario in Eq.~(\ref{E:ex2}) this effect is suppressed due to the deformation in the initial coordinate space.
However, by adding the number of $N_{E}$, the escape of the test nucleons for the first excitation scenario can be diminished.

Based on the obtained $\Delta\langle{\hat{Q}}_{\rm ISQ}\rangle(t)$, the strength function $S(E)$ for the iso-scalar quadruple mode can be calculated through the Fourier transform.
If one excites the nucleus through Eq.~(\ref{E:ex1}), the $S(E)$ can be obtained from
\begin{equation}\label{E:SE}
S(E) = -\frac{1}{\pi\lambda}{\int^{\infty}_{0}}{dt\Delta\langle{\hat{Q}}\rangle(t){\rm sin}\frac{Et}{\hbar}}.
\end{equation}
For the excitation scenario in Eq.~(\ref{E:ex2}), analogous expression can be obtained by replacing the sine in the above equation with the cosine.

\begin{figure}[htbp]
\includegraphics
[width=8.5cm]{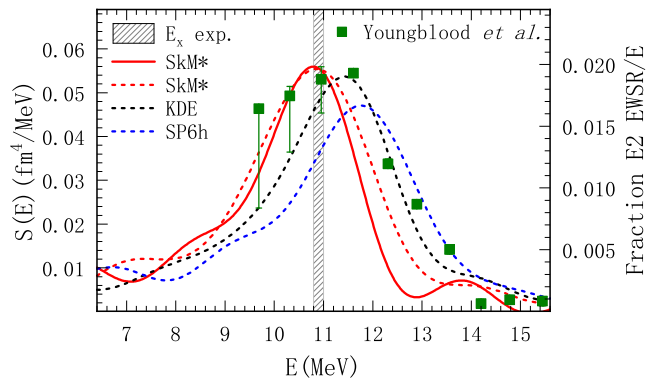}
\caption{
The $S(E)$ of the ISGQR for $\isotope[208]{Pb}$ obtained based on the BUU equation.
The solid line represents the result obtained through the first excitation scenario [Eq.~(\ref{E:ex1})], while the dashed lines correspond to that through the second excitation scenario [Eq.~(\ref{E:ex2})].
The gray hatched band represents the weighted average of the experimental $E_x$ for $\isotope[208]{Pb}$ at $E_{x}$ $=$ $10.9\pm0.1~\rm MeV$.
The green symbols are the experimental E$2$ EWSR by Youngblood {\it et al.}~\cite{YouPRC69}.
} \label{F:SE}
\end{figure}

We exhibit in Fig.~\ref{F:SE} the full BUU calculations of the strength function $S(E)$ for the iso-scalar quadruple excitation of \isotope[208]{Pb} with two standard Skyrme interactions, SkM*, and KDE~\cite{AgrPRC72} and one Skyrme pseudopotential SP$6$h~\cite{WRPRC98}.
For KDE and SP$6$h, we excite the nucleus through Eq.~(\ref{E:ex2}), shown as black and blue dashed lines, respectively. For SkM*, apart from the result through Eq.~(\ref{E:ex2}), represented by the red dashed line, the result through Eq.~(\ref{E:ex1}) is also adopted and the result is shown as the red solid line.
The results with SkM* indicate that the two different excitation scenarios expressed in Eq.~(\ref{E:ex1}) and Eq.~(\ref{E:ex2}) give similar peak energies and width, as well as the shape of the strength function.
In the following, we employ the excitation scenario in Eq.~(\ref{E:ex2}) to generate the iso-scalar quadruple mode of the nucleus, since it is easier to obtain a regular oscillation of $\Delta\langle{\hat{Q}}_{\rm ISQ}\rangle$.
The gray hatched band in Fig.~\ref{F:SE} at $E$ $=$ $10.9\pm0.1~\rm MeV$ represents the weighted average of different experimental values of the excitation energy of ISGQR of \isotope[208]{Pb}~\cite{RocPRC87}.
This value can be reproduced by SkM*, whose $E_x$ $=$ $10.86~\rm MeV$.
Different peak energies of different Skyrme interactions actually reflect their different iso-scalar single particle behavior, i.e., $m_{s,0}^*$.
This correlation can be used to constrain the $m_{s,0}^*$, which is the main topic in the next subsection.
The experimental E$2$ energy-weighted sum rule~(EWSR) measured by Youngblood~{\it et al.}~\cite{YouPRC69} is also included for comparison, i.e., the green symbols in Fig.~\ref{F:SE}.
Since the original data in Ref.~\cite{YouPRC69} are only shown in fraction, the data point are scaled to the peak value of the $S(E)$ obtained with SkM*.
We notice from the figure that its overall shape can be well reproduced within the framework of the BUU equation.

\subsection{Correlation between $E_{x}$ and $m_{s,0}^*$}

Based on the quantal harmonic oscillator approach \cite{Boh1975}, one can obtain a direct relation between the excitation energy $E_{x}$ of the ISGQR and the iso-scalar nucleon effective mass,
\begin{equation}\label{eq.213}
E_{x} = \sqrt{\frac{2m}{m^{*}_{s,0}}}\hbar\omega_{0},
\end{equation}
where $\hbar\omega_{0}$ is the frequency of the harmonic oscillator, which is related to the restoring force of the ISGQR, and $m^{*}_{s,0}$ is the iso-scalar effective mass at the nuclear saturation density $\rho_0$.
By this semiempirical equation, we obtain that a strong linear correlation exists in between $10^{3}/{E_{x}}^2$ and ${m^{*}_{s,0}}/{m}$, \begin{equation}\label{E:E-m}
\frac{10^3}{{E_{x}}^2} = \frac{m^{*}_{s,0}}{m}k+b,
\end{equation}
where the slope $k$ is approximately equal to ${10^3}/{2(\hbar\omega_{0})^2}$ and $b$ is the intercept. 
Based on the BUU equation, we can calculate the $E_x$ of the ISGQR for certain nucleus employing various Skyrme interactions with different $m^*_{s,0}$.

\begin{figure}[htbp]
\includegraphics
[width=8.0cm]{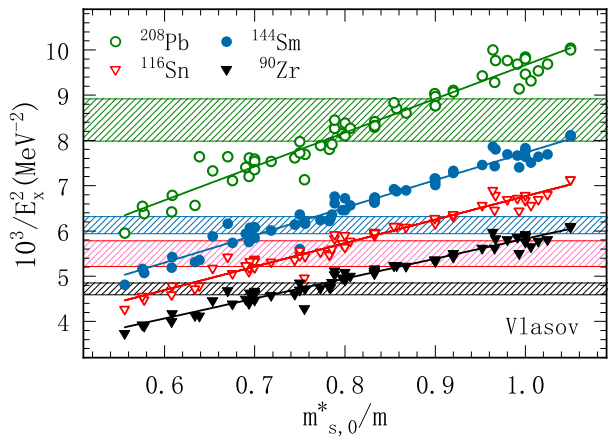}
\caption{
The correlation between the $m^{*}_{s,0}/m$ and the $E_{x}$ of the ISGQR for $\isotope[208]{Pb}$, $\isotope[144]{Sm}$, $\isotope[116]{Sn}$ and $\isotope[90]{Zr}$ obtained by pure Vlasov calculation with 65 Skyrme interactions.
Their linear fittings are shown in solid lines.
The hatched bands correspond to the experimental $E_{x}$ of the ISGQR these nuclei.
} \label{fig.4}
\end{figure}

In order to obtain the correlation between the $E_x$ and the $m^*_{s,0}$, we select $65$ representative Skyrme interactions or Skyrme pseudopotentials, to calculate the strength function of the ISGQR of various heavy nuclei based on the Eq.~(\ref{E:ex2}).
All of these interactions (SV-m56-O, SkI5, SkI3, SGI, SkA, SV-m64-O, SkI6, SkMP, SLy9, SAMi, SLy6,  NRAPR, SkI1, SLy4, SLy0, SLy1, SLy3, SV-mas07, KDE0v, KDE0v1, SKRA, KDE, SK272, Rs, Gs, SGII, SkM, SkM*, SK255, SV-mas08, MSL0, MSL1, SkT7, SkT7a, SkT8, SkT8a, SkT9, SkT9a, SkS3, SkS1, SkS4, SV-sym32, SkO', SV-bas, SV-K241, BSk13, BSk5, SV-min, Skxs20, SKXm, Skxs15, Ska25s20, Ska35s20, SKX, SkSC15, SkT1, SkT2, SkT3, SKXce, Ska35s15, Ska45s20, BSk1, MSk7, details about these standard Skyrme interactions can be found in Ref.~\cite{DutPRC85} and references therein, Sk$\chi$m*~\cite{ZZPLB777}, SP$6$h~\cite{WRPRC98}) can roughly reproduce the empirical values of several characteristic quantities of the nuclear equation of state~\cite{RocPPNP101}, i.e., nuclear saturation density $\rho_{0}$, energy per nucleon for symmetric nuclear matter at saturation density $E_{0}$, the nuclear incompressibility at saturation density $K_{0}$, and the symmetry energy at saturation density $E_{\rm sym}(\rho_0)$.
The $m^{*}_{s,0}/m$ in these Skyrme interactions range from $0.56$ to $1.05$, which can be calculated through Eq.~(\ref{eq.29}).

Figure~\ref{fig.4} displays the pure Vlasov results of the inverse squared excitation energy $1/E_x^2$ of ISGQR versus $m^{*}_{s,0}/m$ with the above $65$ Skyrme interactions for $\isotope[208]{Pb}$, $\isotope[144]{Sm}$, $\isotope[116]{Sn}$ and $\isotope[90]{Zr}$.
The obtained $E_{x}$ of the ISGQR decreases from light to heavy nuclei, which follows the general trend.
As can be noticed in the figure, the $1/E_x^2$ and $m^{*}_{s,0}/m$ for these nuclei exhibit nice linear correlations (their linear fits are shown as solid lines), which indicates that Eq.~(\ref{E:E-m}) holds for a large nuclear mass range.
The experimental $E_{x}$ of the ISGQR for different nuclei extracted in Ref.~\cite{YouPRC69,GuptaPRC97} are also included in Fig.~\ref{fig.4} with different hatched bands, i.e., $10.89\pm0.30~\rm MeV$, $12.78\pm0.20~\rm MeV$, $13.50\pm0.35~\rm MeV$ and $14.56\pm0.20~\rm MeV$ for $\isotope[208]{Pb}$, $\isotope[144]{Sm}$, $\isotope[116]{Sn}$ and $\isotope[90]{Zr}$, respectively.
Due to the limitation of the density functional, the Skyrme interactions which reproduce the experimental $E_x$ of heavier nuclei give slightly lower $E_x$ of lighter nuclei compared with their experimental value, and vice versa, though the deviations are very small.
This result is consistent with those from similar systematic analyses with Skyrme interactions based on the RPA~\cite{BonPRC98} and the macroscopic Langevin equation~\cite{SJPRC101}.

\begin{figure}[htbp]
\includegraphics
[width=8.0cm]{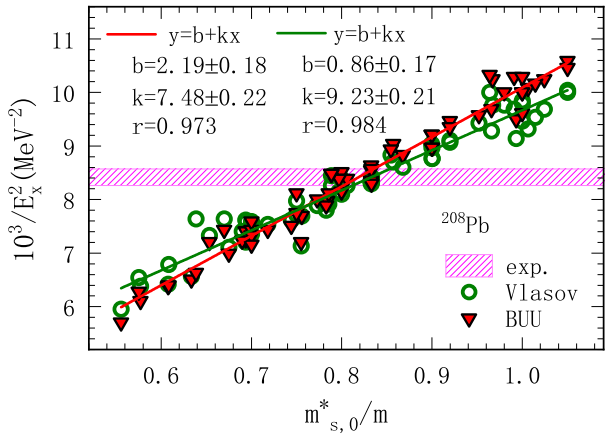}
\caption{
The correlation between the $m^{*}_{s,0}/m$ and the $E_{x}$ of the ISGQR for $\isotope[208]{Pb}$.  The green circles and red triangles are results from the pure Vlasov and the full BUU calculations, respectively, for different Skyrme interactions.
Their linear fittings are shown in solid lines, whose slope $k$, intercept $b$ and Pearson coefficient $r$ are also listed.
The pink hatched band corresponds to the weighted average of the experimental $E_{x}$ $=$ $10.9\pm0.1~\rm MeV$.
} \label{fig.5}
\end{figure}

In principle, mean-field calculations are more suitable for heavier nuclei, therefore we choose \isotope[208]{Pb} to constrain the $m^{*}_{s,0}$.
In Fig.~\ref{fig.5}, in addition to the pure Vlasov result of the $E_{x}$ - $m^{*}_{s,0}/m$ correlation, the result from full BUU calculation with the nucleon-nucleon scatterings is also included.
Through a linear fit, shown as solid lines in Fig.~\ref{fig.5}, we obtain the slope $k$ and intercept $b$ in the relation expressed by Eq.~(\ref{E:E-m}), and they are also given in Fig.~\ref{fig.5} along with the Pearson coefficient of the fittings.
As shown in the figure, the results of Vlasov and BUU calculation are quite similar except a small deviation. 
Due to the damping provided by the nucleon-nucleon scatterings, the slope $k$ obtained from BUU calculations is a little bit larger then that in the Vlasov calculation.
The constrains on the ${m^{*}_{s,0}}/{m}$ can be obtained through ${m^{*}_{s,0}}/{m}$ $=$ $(10^3/E_x^2 - b)/k$.
Here a weighted average value of $E_x$ $=$ $10.9\pm0.1~\rm MeV$ over several experiment measurements~\cite{RocPRC87} is adopted, and represented by pink hatched band in Fig.~\ref{fig.5}.
For the Vlasov calculation, we have
\begin{equation}\label{E:ms1}
\frac{m^{*}_{s,0}}{m} = 0.83 \pm 0.04,
\end{equation}
while for the BUU calculation
\begin{equation}\label{E:ms2}
\frac{m^{*}_{s,0}}{m} = 0.82 \pm 0.03.
\end{equation}
The uncertainties in the above constraints on $m_{s,0}^*/m$ are propagated from the uncertainties of $k$, $b$ and $E_x$.
The central value of the constraint from the BUU calculation is slightly lower than that from the Vlasov calculation, whereas they are within the errors of each other.
This indicates the inclusion of the nucleon-nucleon scatterings ($2$p-$2$h correlation) has only negligible effect on the constraint of the $m_{s,0}^*/m$.
These effective mass constraints are consistent with the $m^{*}_{s,0}/m\approx0.8-0.9$ obtained from the ISGQR of the spherical nucleus $\isotope[144]{Sm}$ and the deformed nucleus $\isotope[154]{Sm}$ \cite{YosPRC88}.
The constraint on $m_{s,0}^*$ from the Vlasov calculation is also in agreement with the constraint $m_{s,0}^*/m$ $=$ $0.91 \pm 0.05$ obtained from a conventional analysis~\cite{ZZPRC93} and $m_{s_0}^*/m=0.87 \pm 0.04$ from a recent Bayesian analysis~\cite{ZZCPC45} of ISGQR of \isotope[208]{Pb} based on the Skyrme-RPA calculations, where the $2$p-$2$h correlation is also missing.

\section{Summary}\label{4}
In the present study, the nuclear ISGQR has been investigated based on the BUU transport equation.
The mean-field part of the BUU equation is described by the Skyrme nucleon-nucleon effective interaction, and numerically solved through a lattice Hamiltonian method.
We employ the stochastic approach to deal with the collision term of the BUU equation, which provides a two-body dissipation and is responsible for the collisional damping of the nuclear collective motions.
The measured ISGQR width and the overall shape have been reproduced based on the BUU equation with a proper in-medium nucleon-nucleon cross section.
Unlike in the case of the IVGDR, where collisional damping is only responsible for about half of the total width, we find that the width of ISGQR of heavy nuclei (\isotope[208]{Pb}) is dominantly exhausted by the collisional damping, which suggests the ISGQR to be an ideal site for constraining the in-medium nucleon-nucleon cross section.
A strong linear correlation between the iso-scalar nucleon effective mass $m^{*}_{s,0}$ and $1/E_x^2$, the inverse of the squared excitation energy of ISGQR, has been obtained for \isotope[90]{Zr}, \isotope[116]{Sn}, \isotope[144]{Sm} and \isotope[208]{Pb} based on the collisionless BUU equation, i.e., the Vlasov equation.
Through the weighted average of the experimental excitation energy $E_x$ of \isotope[208]{Pb}, we have extracted a constraint on the iso-scalar nucleon effective mass at $m^{*}_{s,0}/m$ $=$ $0.83 \pm 0.04$.
If we include the collision term of the BUU equation, which embodies the beyond mean-field $2$p-$2$h correlation, the constraint on $m^{*}_{s,0}/m$ only decreases slightly to $0.82 \pm 0.03$, which indicates that the $2$p-$2$h correlation has negligible effect on the excitation energy, and subsequently on the constraint of the iso-scalar nucleon effective mass.

\begin{acknowledgments}
We thank Chen Zhong for setting up and maintaining the GPU server.
This work is partially supported by the National Natural Science Foundation of China under Contracts No. $11890714$, No. $11891070$ and No. $11905302$, the Guangdong Major Project of Basic and Applied Basic Research No. $2020$B$0301030008$,  and the Key Research Program of the CAS under Grant No. No. XDB34000000.
\end{acknowledgments}
\bibliography{GQR-ms-BUU}
\end{CJK*}
\end{document}